\documentclass{ws-procs975x65}

\begin{document}

\title{EXISTENCE OF HORIZONS IN ROBINSON--TRAUTMAN SPACETIMES OF ARBITRARY DIMENSION}

\author{OTAKAR SV\'{I}TEK}

\address{Institute of Theoretical Physics, Charles University in Prague, Faculty of Mathematics and Physics, V~Hole\v{s}ovi\v{c}k\'ach 2, 180~00 Praha 8, Czech Republic\\
ota@matfyz.cz}

\begin{abstract}
We derive the higher dimensional generalization of Penrose--Tod equation describing past horizon in Robinson--Trautman spacetimes with a cosmological constant and pure radiation. Results for $D=4$ dimensions are summarized. Existence of its solutions in $D>4$ dimensions is proved using tools for nonlinear elliptic partial differential equations.
\end{abstract}

%\submitto{MGM 09 Proceedings}
%\pacs{04.20.Gz, 04.50.Gh}

\bodymatter

\section{Introduction}
Our concern here is to locate the past (white hole) horizon. In general dynamical situations this might be rather nontrivial since the obvious candidate - event horizon - is a global characteristic and therefore the full spacetime evolution is necessary in order to localize it. Therefore, over the past years different quasi-local characterizations of black hole boundary were developed. The most important ones being apparent horizon \cite{hawking-ellis}, trapping horizon \cite{hayward} and isolated or dynamical horizon \cite{ashtekar}. The basic {\it local} condition in the above mentioned horizon definitions is effectively the same: these horizons are sliced by marginally trapped hypersurfaces with vanishing expansion of outgoing (ingoing) null congruence orthogonal to the surface.

For the vacuum four dimensional Robinson--Trautman solutions without cosmological constant the location of the horizon together with its general existence and uniqueness has been studied by Tod \cite{tod}, and further by Chow and Lun \cite{chow-lun}. These results were recently extended to nonvanishing cosmological constant \cite{PodSvi:2009}. 

Robinson--Trautman spacetimes (containing aligned pure radiation and a cosmological constant $\Lambda$) in any dimension were obtained \cite{podolsky-ortaggio} by requiring the existence of a twistfree, shearfree and expanding null geodesic congruence. They have arrived at the following metric valid in higher dimensions
\begin{equation}
 d s^2=\frac{r^2}{P^2}\,\gamma_{ij}\,d x^i d x^j-2\,d u d r-2H\, d u^2 
\end{equation}
where $2H=\frac{{\cal R}}{(D-2)(D-3)}-2\,r(\ln P)_{,u}-\frac{2\Lambda}{(D-2)(D-1)}\,r^2-\frac{\mu(u)}{r^{D-3}}$. The unimodular spatial $(D-2)$-dimensional metric $\gamma_{ij}(x)$ and the function $P(x,u)$ must satisfy the field equation ${\cal R}_{ij}=\frac{{\cal R}}{D-2}h_{ij}$ (with $h_{ij}=P^{-2}\gamma_{ij}$ being the rescaled metric and ${\cal R}_{ij}$ its Ricci tensor), $\mu(u)$ is a ``mass function'' (we assume $\mu>0$).

\section{Past horizon}
To localize the past horizon, we will be dealing only with the condition of vanishing expansion defining the marginally trapped hypersurfaces. The explicit parametrization of the {\em past horizon} hypersurface is $r=R(u,x^{i})$ such that its intersection with each $u=u_{1}$ slice is an outer marginally past trapped ($D-2$)-surface.

By straight-forward computation (using a generalization of the tetrad formalism to arbitrary dimension) one easily calculates the expansion associated with the outgoing null congruence $l^{a}$ (orthogonal to $r=R(u=const,x^{i})$ surface) to be $\Theta_{l}= \frac{D-2}{r}$ meaning that it is diverging. This is exactly what one assumes when dealing with the past trapped surface and is the additional condition in the definition of trapping horizon \cite{hayward}.

Expansion of the ingoing null congruence $n^{a}$ can be calculated using the formula $\Theta_{n}=n_{a;b}\,p^{ab}$, where the tensor $p^{ab}=g^{ab}+2\,l^{(a}n^{b)}$ corresponds to the hypersurface projector. From $\Theta_{n}= 0$  we get the marginally trapped hypersurface condition (equivalent to Penrose--Tod equation in four dimensions)
\begin{equation*}
{\cal R}-{\frac{2(D-3)}{D-1}}\Lambda R^{2}-{(D-2)(D-3)}\frac{\mu}{R^{D-3}}-{2(D-3)}\Delta(\ln R)-
\end{equation*}
\begin{equation}\label{PT}
-{(D-4)(D-3)}(\nabla \ln R)\cdot (\nabla \ln R)= 0
\end{equation}
It is a nonlinear second order partial differential equation, where both the Laplacian and scalar product in the last term correspond to the Einstein metric $h_{ij}$. Interesting property of this equation is that for $D>4$ its nonlinearity is much worse since the term quadratic in derivatives appears.

\section{Results for $D=4$}
The uniqueness and existence results for equation \eqref{PT} are derived in Ref.~\citen{PodSvi:2009} (with $\mu=2m=const.$ and using the previous works Refs.~\citen{tod,chow-lun}) and the results are summarized in the following table:
\vspace{-0.4cm}
\begin{table}[h]
$$
\begin{array}{||c||c|c|c||}
\hline \hline
\mbox{RESULTS} & \Lambda=0 & \Lambda<0 & \Lambda>0 \\
\hline \hline
\mbox{Existence} & \mbox{Always} & \mbox{Always} & \Lambda < \frac{4}{9\mu^2} \\
\hline
\mbox{Uniqueness} &  \mbox{Always} & \mbox{Always}  & R < \sqrt[3]{\frac{3\mu}{2\Lambda}}\\
\hline \hline
\end{array}
$$
\end{table}
\vspace{-1cm}

\section{$D>4$ : Existence of the solution}
The methods used in $D=4$ are not applicable when the equation is of the form (after the substitution $R=C e^{-u}$ in (\ref{PT}), assuming $u\geq 0$ with a suitable constant $C$)
\begin{equation}
\Delta u=F(x,u,\nabla u)\ ,
\end{equation}
where $F$ is quadratic in gradient.

To prove existence of the solution to this quasilinear equation one can proceed by combining several steps (motivated by Ref.~\citen{Kuo} and using results from Refs.~\citen{Besse,Boccardo,Gilbarg-Trudinger}). These steps include the use of Maximum Principle, Fredholm alternative, truncature of functions, Schauder Fixed Point theorem and elliptic estimates. The main requirement one has to satisfy is the existence of sub- and super-solutions.

One can construct constant sub- and super-solutions (assuming $u_{min}>0$  and ${\cal R}>0$) for any $\Lambda\leq 0$, but for positive cosmological constant one has to demand 
\begin{equation}\label{restriction}
\frac{2 {\cal R}}{(D-1)(D-2)(D-3)\mu}(\frac{{\cal R}}{2 \Lambda})^{\frac{D-3}{2}}>1\ .
\end{equation}
Interestingly, this last condition reduces in the four-dimensional case (which can be included) to the condition from the above table for the existence of the solution when $\Lambda > 0$ (${\cal R}$ asymptotically approaches $2$).

Since according to mathematical results any manifold (including compact ones) of dimension greater than or equal to $3$ can be endowed with a complete Riemannian metric of constant negative scalar curvature \cite{Aubin,Lohkamp} one should also consider that ${\cal R}<0$ for our $D-2$-dimensional spatial hypersurface. However, one can construct constant sub- and super-solutions only for negative cosmological constant.

\section{Conclusion}
We have derived the generalization of the Penrose--Tod equation to higher dimensional Robinson-Trautman spacetimes including cosmological constant and pure radiation. Using several mathematical tools we have proved the existence of its solution for any $\Lambda\leq 0$ and for ${\cal R}>0$. The limitations arising for positive $\Lambda$ are naturally related to the more complicated horizon structure of relevant spacetimes. 

\section*{Acknowledgements}
This work was supported by grants GACR 202/07/P284, GACR 202/09/0772 and the Czech Ministry of Education project Center of Theoretical Astrophysics LC06014.

\end{document}